\documentclass{iopjournal}
\usepackage{amsmath}
\usepackage{amsfonts}
\usepackage{amssymb}
\usepackage{booktabs} 
\usepackage{multirow} 

\begin{document}

\articletype{Paper} 

\title{Thermal Radiosensitization Beyond Misrepair: A Mechanistic Model of Temperature-Enhanced DNA Vulnerability}

\author{
	Jos\'e L. Rodr\'iguez-Amado$^1$	, 
	Edwin Mun\'evar$^1$\footnote{ORCID: \url{https://orcid.org/0000-0002-0578-7717}},
	C\'esar A. Herre\~no-Fierro$^1$\footnote{ORCID: \url{https://orcid.org/0000-0003-2394-4322}}
	 and 
	Adriana M. De Mendoza$^{2,*}$\footnote{ORCID: \url{https://orcid.org/0000-0002-8006-1756}}
	}

\affil{$^1$Facultad de Ciencias y Educaci\'on, Universidad Distrital Francisco Jos\'e de Caldas,  Bogot\'a D.C., Colombia}

\affil{$^2$Physics department, Pontificia Univsersidad Javeriana, Bogot\'a D.C., Colombia}

\affil{$^*$Author to whom any correspondence should be addressed.}

\email{a.demendoza@javeriana.edu.co}

\keywords{Thermal enhancement of radiation, Hyperthermia, Radiotherapy, DNA repair inhibition, Radiosensitization, DNA breathing, Mechanistic Mathematical modeling.}

\begin{abstract}
\textit{Objective:} Hyperthermia (HT), characterized by elevated tissue temperatures above physiological levels, is a well-established radiosensitizer. When combined with radiotherapy (RT), forming thermoradiotherapy (TRT), a synergistic effect is observed across \textit{in vitro}, \textit{in vivo}, and clinical studies. The greatest radiosensitization occurs when HT and RT are applied simultaneously. This work aims to explore physical mechanisms—beyond DNA repair inhibition—that contribute to this synergy.\\

\textit{Approach:} We developed a biophysical model for the thermal enhancement ratio (TER), incorporating temperature-dependent variations in the number of vulnerable DNA sites, the DNA–ion/particle interaction cross-section, and other physicochemical parameters. These include ion production rate, diffusion processes, and medium density. The model includes misrepair effects phenomenologically, that make it consistent with other studies.\\

\textit{Main results:} The model reproduces TER values observed under simultaneous HT and RT in isolated plasmids with variable temperature. Our results indicate that, in addition to misrepair, other physical factors contribute to radiosensitization under concurrent treatment. Among these, the temperature-dependent amplification of DNA–ion/particle interaction cross-section—driven by enhanced DNA thermal fluctuations structure—emerges as the second most influential factor.\\

\textit{Significance:} These findings suggest that thermal radiosensitization arises not only from impaired repair, but also from increased physical vulnerability of the DNA. The model provides mechanistic insight for optimizing TRT parameters.
\end{abstract}

\vspace{0.4cm}

\section{Introduction}

Hyperthermia treatment (HT), defined as the controlled elevation of tissue temperature above physiological levels, is widely recognized as a potent radiosensitizer \cite{Overgaard, Horsman}. Its radiosensitizing effects have been demonstrated in multiple \textit{in vitro}, \textit{in vivo}, and clinical studies \cite{Overgaard2013, Overgaard, Mei, Elming}. When combined with radiotherapy (RT) --a modality known as thermoradiotherapy (TRT)-- significant radiation dose reductions, up to an order of magnitude, have been achieved. The extent of radiosensitization depends on factors such as temperature, treatment duration, sequence and timing of application, and the biological system under study \cite{Overgaard2013, Overgaard, van_Leeuwen2017, van_Leeuwen20172}. The synergistic effect of TRT is typically quantified by the thermal enhancement ratio (TER), defined as the ratio between the biological effect of radiation treatments with and without heat sensitization. The highest TER values are observed when HT and RT are applied simultaneously, with radiosensitization decaying exponentially as the time interval between treatments increases. Moreover, the treatment order matters: a faster TER decay occurs when RT precedes HT \cite{Overgaard2013, Overgaard, Lindegaard3}. \\
\vspace{0.5cm}

Multiple hypotheses have been proposed to explain the synergistic interaction between hyperthermia and radiation therapy~\cite{Oei,Lepock,Lepock1,Lepock2,Roti,Hildebrandt}, with the most widely accepted being the heat-induced inactivation of DNA repair proteins, commonly known as the  \textit{misrepair} mechanism~\cite{Oei}. In our previous work, we suggested that radiosensitization arises from increased DNA vulnerability due to heat, particularly through the thermal denaturation of DNA repair proteins~\cite{DeMendoza,DeMendoza1}. However, since the renaturation times of these proteins are often comparable to or longer than the observed timescale of TER decay~\cite{Jungbauer,Parsell,Seckler}, it remains uncertain whether misrepair alone can fully account for the radiosensitization effect. This mechanism may operate regardless of whether hyperthermia is applied before, during, or after radiation. Notably, the pronounced TER peak observed under simultaneous application of HT and RT strongly suggests the contribution of additional temperature-dependent mechanisms that act on much faster timescales. \\
\vspace{0.5cm}

Several experimental and theoretical studies offer valuable, though still loosely connected, insights into how temperature modulates radiosensitivity. Experimentally, radiation-induced cell mortality has been strongly correlated with the formation of DNA double-strand breaks (DSBs)~\cite{Belli, Taucher}, a relationship captured by the statistical model of Chadwick and Leenhouts~\cite{Chadwick}. Tomita \textit{et al.} demonstrated that DSBs in irradiated plasmids increase with temperature~\cite{Tomita}. On the theoretical side, Deppman \textit{et al.} quantified DSB formation based on physical parameters such as DNA cross-section, density, and energy deposition~\cite{Deppman2004}. More recently, Ramos-Méndez \textit{et al.} integrated temperature-dependent reaction rates and diffusion coefficients into TOPAS-nBio simulations to reproduce Tomita’s data~\cite{Ramos}. However, their model predicts a linear increase in break yield with temperature, failing to capture the experimentally observed exponential trends, especially for single-strand breaks (SSBs). From the perspective of DNA biophysics, the statistical mechanics model of Peyrard and Bishop~\cite{Peyrard1,Dauxois1993} shows that local strand separation (\textit{DNA breathing}) increases exponentially with temperature, independent of enzymatic activity. This suggests that rising temperature alone can amplify DNA vulnerability, providing a purely physical pathway for radiosensitization distinct from repair inhibition.\\
\vspace{0.5cm}

Motivated by these findings, the present study develops a mechanistic model of the thermal enhancement ratio (TER) in thermoradiotherapy. We identify several temperature-dependent factors: (i) the number of ions created by water radiolysis and their diffusion, (ii) the number of vulnerable DNA sites, (iii) the probability of collisions between DNA and damaging agents (e.g., ions, reactive oxygen species, photons, electrons, neutrons, protons, $\alpha$- and $\beta$-particles), and (iv) the water density.  Our analysis shows that, after DNA repair inhibition, the most significant contributor to TER is the increase in DNA collision cross-section driven by \textit{DNA breathing}. This thermally activated fluctuation enhances the likelihood of radiation-induced strand breaks, particularly under simultaneous HT and RT. By quantifying the relative contributions of each factor across the therapeutic hyperthermia range (40–50$^\circ$C), our model offers a physically grounded explanation for the peak in radiosensitization observed under concurrent treatment, and reconciles experimental observations with theoretical predictions.\\
\vspace{0.5cm}

This article is organized as follows: Section~\ref{DSBp} derives the relationship between DNA rupture probability and physical parameters such as linear energy transfer (LET), target molecular density, and collision cross-section, revisiting the linear-quadratic (LQ) model in this context. Section~\ref{ters} presents the proposed temperature-dependent expression for the thermal enhancement ratio. In Section~\ref{cs}, we recall the Peyrard-Bishop model to describe the exponential temperature dependence of the DNA cross-section. 

\section{Methods}

In order to obtain a thermal enhancement function expressed in terms of the physical and biological parameters that may vary with HT treatment time and temperature, we follow the procedure of Deppman \textit{et al.}, to reconstruct the dependency of DNA-ruptures with physical parameters such as LET, target molecular density, and cross section \cite{Deppman2004}. Later on, we recall the quantitative relationship between DNA rupture and cell survival probability from the \textit{linear-quadratic model} (LQ-model) \cite{Chadwick}. Finally, we derive an expression for the thermal enhancement ratio (TER), as a function of the aforementioned parameters, and explore the temperature dependency of the expression. To facilitate the reading of this section,  all the variables and parameters of the model are described in Table \ref{model_parameters}.\\

\begin{table}[ht]
	\centering
	\caption{Model parameters and variables. \label{model_parameters}}
	\begin{tabular}{cll}
		\hline
		\textbf{\textcolor{white}{---}Symbol}\textcolor{white}{---} & \textbf{\textcolor{white}{----------------------------}Definition} & \textbf{Units} \\
		\hline
		$T$ & Temperature & K \\
		$t$ & Hyperthermia treatment duration & min \\
		$T_0$ & Physiological reference temperature & K \\
		$T_g$ & Average melting temperature of DNA-repair proteins & K \\
		$a$ & Temperature coefficient (DNA breathing) & K$^{-1}$ \\
		$b$ & Temperature coefficient (repair inhibition) & K$^{-1}$ \\
		$c$ & Amplitude of repair inhibition function & -- \\
		$m$ & Amplitude of the approximated cross-section model & eV/\AA$^2$ \\
		$k\epsilon$ & Number of ions generated per unit track length & m$^{-1}$ \\
		$\sigma$ & Collision cross-section of DNA with ionizing species & m$^2$ \\
		$\sigma_d$ & Collision cross-section of DNA with ions & m$^2$ \\
		$\rho$ & Medium density & kg/m$^3$ \\
		$r_i$ & Diffusion distance of reactive species & m \\
		$n_m$ or $n_i$ & Number of target molecules or vulnerable sites & -- \\
		$f_i$ & Fraction of unrepaired vulnerable sites & -- \\
		$A$ & Pre-exponential factor in Arrhenius law & s$^{-1}$ \\
		$E_a$ & Activation energy & eV \\
		$k_B$ & Boltzmann constant & eV/K \\
		$\mu$ & Mobility of ions or molecules & m$^2$/Vs \\
		$\mathcal{D}$ & Diffusion coefficient & m$^2$/s \\
		$dV$ & Differential volume element & m$^3$ \\
		$dS$ & Differential surface element & m$^2$ \\
		$\zeta$ & Coincidence factor (recombination scale) & m \\
		$D$ & Radiation dose & Gy \\
		$\Delta$ & Fraction of dose contributing to single-hit DSBs & -- \\
		$\xi$ & Probability of two SSBs forming a DSB & -- \\
		$\alpha$ & Linear coefficient of cell kill (LQ model) & Gy$^{-1}$ \\
		$\beta$ & Quadratic coefficient of cell kill (LQ model) & Gy$^{-2}$ \\
		$K$ & Average DNA breakage rate & s$^{-1}$ \\
		\hline
	\end{tabular}
\end{table}

\subsection{Radiation interaction with DNA and rupture probability}\label{DSBp}

\begin{figure}
	\center
	\includegraphics[width=0.7\textwidth]{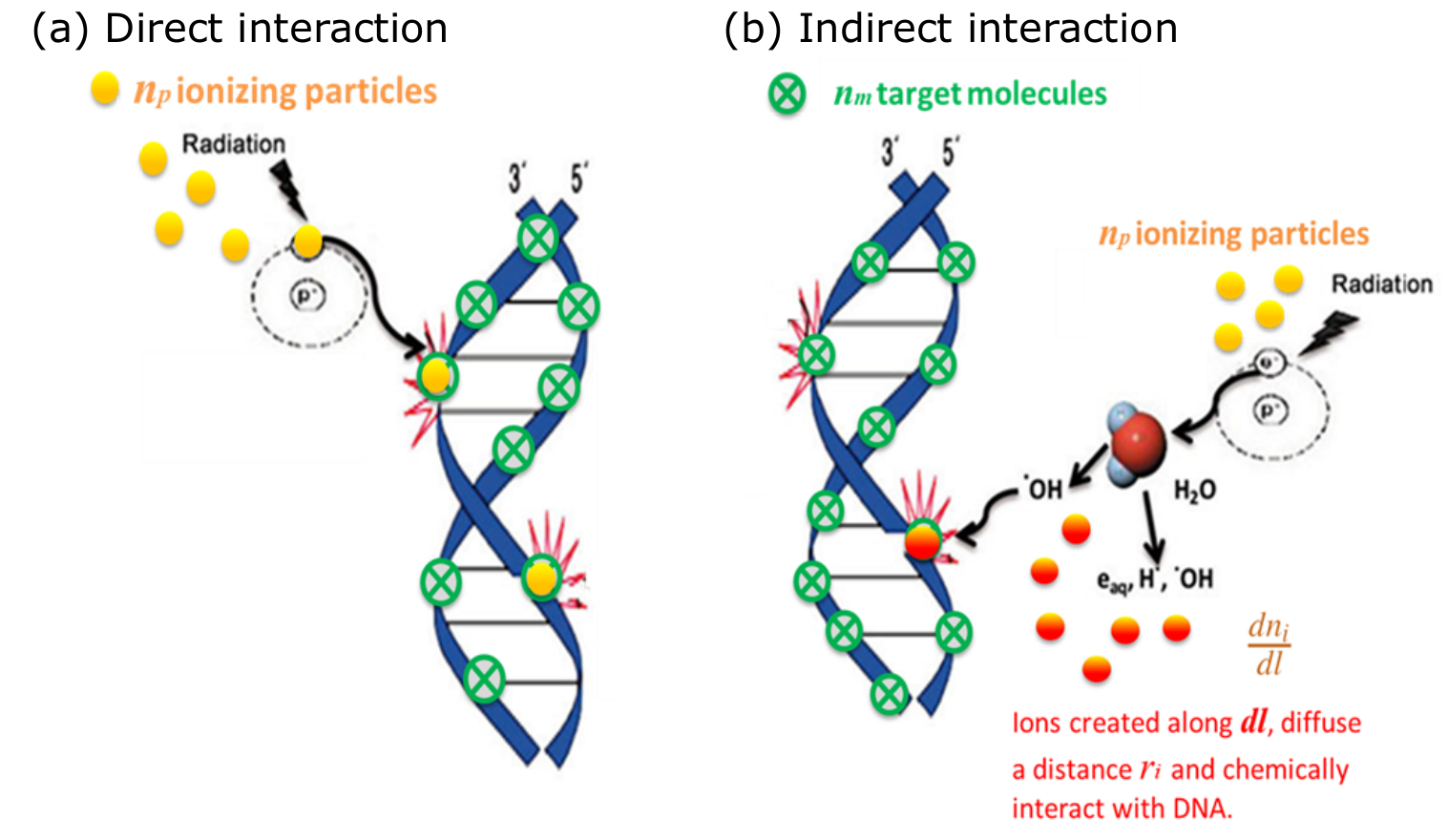}
	\caption{Schematic representation of the mechanisms by which ionizing radiation induces DNA strand breaks. (a) Direct action: Ionizing particles ($n_p$) interact directly with DNA molecules, depositing energy that leads to strand breaks at specific sites ($n_m$ target molecules). (b) Indirect action: Radiation ionizes water molecules near the DNA, producing reactive oxygen species (ROS). These ions diffuse over a distance $r_i$ from the radiation track and interact chemically with the DNA, inducing damage. They cover a volume $dV=dr_idS$, where $dS=2\pi r_idl$. The number of interacting ions per unit track length is denoted by $\frac{dn_i}{dl}$. Both pathways contribute to radiation-induced damage, but their relative importance depends on factors such as radiation quality (LET), DNA structure, and the physicochemical properties of the surrounding medium. \label{esq}}
\end{figure}

Deppman and collaborators \cite{Deppman2004} linked DNA-rupture probability to physical parameters such as density, cross section of collisions with DNA, and LET. To establish this relationship, it must be first considered that DNA rupture occurs after direct or indirect interaction of the incident ionizing particle and the DNA molecule. The contribution of each kind of interaction depends on the type of radiation, being the indirect mechanism dominant for low LET radiation \cite{RBbook}. In the direct path, the particles of the radiation field induce DNA-breaks, while in the indirect mechanism, the radiation particles ionize molecules different than DNA, typically water, in a process called radiolysis. Then, the ions created along a track of length $dl$, diffuse a distance $r_i \sim$180 nm and chemically interact with DNA, leading to DNA damage. These processes are schematized in Figure \ref{esq}.\\
\vspace{0.5cm}

The number of DNA damages $N_d$ induced by direct ionization is  proportional to the number of ionizing particles $n_p$ and the number of DNA target molecules $n_m=\frac{dn_m}{ds}\sigma$, where $\frac{dn_m}{ds}$ is the number of target molecules impacting the incident particles per unit area, and $\sigma$ is the cross section of the collision between particles and target molecules. Therefore, we can write

\begin{equation}\label{dN}
	dN_d=n_p\frac{dn_m}{ds}\sigma.
\end{equation}

\noindent If $\rho_m=dn_m/dV$ is the density of target molecules, Eq. \ref{dN} becomes 

\begin{equation}\label{dN2}
	dN_d=n_p\sigma\rho_mdl.
\end{equation}

To determine the number of DNA damages caused by the indirect mechanism, an additional step must be included: the collision between ions (generated by radiolysis) and target molecules, specifically DNA. Let $\frac{dn_i}{dl}$ represent the number of ions created per unit length by the interaction with a single particle in the radiation field. The linear energy transfer, denoted by $\varepsilon$, is defined as the total energy deposited per unit length,  $\varepsilon=\epsilon_a \frac{dn_i}{dl}$, where $\epsilon_a$ is the energy transferred to each ion.  Defining $k=\frac{1}{\epsilon_a}$, we can write 

\begin{equation}\label{dni}
	dn_i=k\varepsilon dl.
\end{equation}

\noindent Now, let $n_d$ be the number of DNA-damages induced by the $dn_i$ ions, which in turn, were generated by each radiation particle. Therefore, the total number of DNA damages $N_d$ induced indirectly by the action of the radiation field is

\begin{equation}\label{dN3}
	dN_d=n_p dn_d.
\end{equation}

Similarly to how Eq. \ref{dN} was formulated, the number of damages induced by the ions can be written as

\begin{equation}\label{dn}
	dn_d=dn_i \frac{dn_m}{ds}\sigma_d \approx dn_i \rho_m r_i\sigma_d,
\end{equation}

\noindent where $\sigma_d$ is the cross section of the interaction between ions and DNA, and $r_i$ is the mean radius covered by the ions during diffusion ($r_i\sim 180$ nm in water at 37 °C). We reserve $\sigma$ for direct radiation-DNA collisions and $\sigma_d$ for ion-DNA collisions. By substituting Eq. \ref{dni} into Eq. \ref{dn}, and comparing it with the result from the direct pathway in Eq. \ref{dN2}, a relationship between the direct and indirect cross sections can be established:

\begin{equation}\label{sigmas}
	\sigma=\gamma\varepsilon,
\end{equation}

\noindent where $\gamma=\sigma_d k r_i$.\\

It should be noted that expression \ref{dni} is just an approximation because it does not consider the depletion of ions due to recombination. The number of recombination events per unit length, $\frac{dn_r}{dl}$, can be assumed to be proportional to the square of the number of ions, modulated by a coincidence factor $\zeta$ with units of length. Accordingly, the corrected number of available ions is given by:

\begin{equation}\label{rec}
	\nonumber dn'_i=dn_i-dn_r=k\varepsilon dl (1-\zeta k\varepsilon) \simeq \frac{k\varepsilon dl}{1+\zeta k\varepsilon}.
\end{equation}

\noindent In the last expression of Eq.~\ref{rec}, we have used the geometric series expansion $(1 - x)^{-1} = \sum_{n=0}^\infty x^n$ under the assumption of low LET, i.e., small $\zeta k \varepsilon$. Substituting this correction into Eq.~\ref{dn} leads to a modified expression for the cross section $\sigma$ in Eq.~\ref{sigmas}, which successfully reproduces the sigmoidal behavior observed in various experimental studies~\cite{Deppman2004,Scholz}:

\begin{equation}\label{sigmas2}
	\sigma=\frac{\gamma\varepsilon}{1+\zeta k\varepsilon}.
\end{equation}

To determine the average number of DNA-breaks, let us consider again Eq. \ref{dN3} which now gets

\begin{align*}
	dN_d&=n_p (k\varepsilon dl)\frac{dn_m}{ds}\sigma_d =n_p (k\varepsilon dl) \frac{dn_m}{dV}\sigma_d r_i\\
	&= (n_p \varepsilon dl) \rho_m \left( \frac{\gamma}{1+\zeta k \varepsilon}\right)\\
	&= dD \rho_0 n_m \frac{ \gamma}{1+\zeta k \varepsilon},
\end{align*}

\noindent where $\rho_0$ is the mass density of the sample contained in the volume  $dV$ (mainly water), and we have used $V=n_m/\rho_m=m/\rho_0$. Introducing $dD= n_p \varepsilon dl/m$ as the differential amount of radiation dose, the increment in the number of damages, which is equal to the reduction of the target molecules can be expressed as

\begin{equation}\label{N5}
	\frac{dN_d}{dD}=-\frac{dn_m}{dD}=n_m\frac{\gamma  \rho_0}{1+\zeta k\varepsilon}.
\end{equation}

From the right hand side of Eq.\ref{N5}, it comes that 

\begin{equation}\label{km}
	n_m=n_{0m}e^{-K_m D},
\end{equation}

\noindent where $n_{0m}$ is the initial number of target molecules, and $K_m$ is the probability of DNA-breaks per unit of dose 

\begin{equation}\label{alpha}
	K_m=\frac{\gamma \rho_0}{1+\zeta k\varepsilon}=\frac{\sigma_d k  \rho_0 r_i}{1+\zeta k\varepsilon}.
\end{equation}

\noindent The denominator in Eq.\ref{alpha} is important only when recombinations are significant.\\

In the low dose regime, and considering that just a fraction $f_0$ of DNA-breaks is not repaired, the amount of average DNA-breaks $\alpha=K_mn_{0m}$  needs to be corrected as follows: 

\begin{equation}\label{alpha2}
	\alpha=f_0\frac{n_{0m}\sigma_d k \rho_0 r_i}{1+\zeta k\varepsilon}.
\end{equation}

\vspace{0.4cm}
It is important to note that the derived result applies to DNA strand breaks --whether single- or double-stranded-- when induced by a single ion or particle. However, double-strand breaks (DSBs) can also arise via the two-hit mechanism, in which two spatially proximate single-strand breaks (SSBs) occur on opposite strands  (see Figure \ref{DNAb}(a) for illustration). In this framework, we define $\alpha$ as the average number of single-hit DSBs and $\beta$ as the average number of two-hit DSBs:

\begin{align}
\nonumber \alpha&=f_0\Delta \: n_{0m}\frac{ \sigma_d k \rho_0 r_i}{1+\zeta k\varepsilon},\\
	\label{beta3} \beta&=\xi \: f_1f_2n_{01m}n_{02m}\left((1-\Delta) \frac{\sigma_d k \rho_0 r_i}{1+\zeta k\varepsilon}\right)^2.
\end{align}

\noindent Again, $\Delta$ is the fraction of the total dose invested into the $1-hit$ mechanism, and the remaining fraction $(1 - \Delta)$ is allocated to the $\beta$ pathway. Recall that $\beta$ is proportional to the square of $\alpha$ times a spatial coincidence factor $\xi$.

\subsection{The LQ model}\label{LQm}

Double strand breaks are strongly correlated with cell death or loss of its proliferative capacity \cite{Radford,Fu}. The unrepaired DSBs lead to chromosomal aberrations, which in many cases end up in mitotic catastrophe (i.e, cell death), or in check points that do not progress to mitosis (cell cycle arrest) \cite{RBbook}. In the context of radiobiology, both cases are considered as clonogenic cell death.\\
\vspace{0.5cm}

The linear-quadratic (LQ) model is a foundational tool in radiobiology that characterizes the relationship between radiation dose and clonogenic cell survival. Widely applied in radiotherapy planning, it accounts for both linear and quadratic components of radiation-induced damage, thereby enabling predictions of treatment efficacy across varying dose fractionation schemes \cite{McMahon,Wang}. Although initially introduced as an empirical fit to experimental data \cite{Lea1942}, a mechanistic interpretation was later provided by Chadwick and Leenhouts  \cite{Chadwick}, who established a mathematical link between the formation of DNA double-strand breaks (DSBs) and cell survival probability. To compute cellular survival probability as a function of the average number of DNA-DSBs, we shall describe the process followed by Chadwick and Leenhouts, which starts from three main assumptions: (1) The target of the radiation is the DNA double helix, (2) the number of DSBs is a function of the radiation dose, and (3) the cell is able to repair, at least partially, the DNA damage induced by radiation. In this model, $n$ is the number of critical bonds per unit mass, and $K$ the probability per unit of radiation dose $D$ of bond breaking. The change in the number of critical bonds $dn/dD$ is proportional to $n$ and decays with the rate $K$ as follows: 

\begin{equation}
	\frac{dn}{dD}=-K n, 
	\label{k}
\end{equation}

\noindent so that $n=n_0e^{-KD}$. Here $n_0$ is the original number of critical bonds at $D=0$. Thus, the average number of broken bonds reads

\begin{equation}
	N_d=n_0-n=n_0(1-e^{-KD}).
\end{equation}

Defining $f_0$ as the probability that the broken bond will not be repaired after damaged, the average number of lethal strand-breaks is now

\begin{equation}
	N_d=f_0n_0(1-e^{-KD}).
\end{equation}

\noindent For low radiation doses, the mean number of DSBs per unit of dose $N_d/D$ is given by $n_0f_0K$. \\
\vspace{0.5cm}

As mentioned before, DSBs can be obtained through two types of events. Breaking the double helix with one collision or ``hit" of the radiation field (e.g. a photon, particle or ion), or by means of two SSBs, which are near enough to accomplish DNA disruption, i.e. ``2-hits". The mean number of DSBs (per radiation dose D), which is produced by 1-hit is given by $\alpha=\Delta n_0f_0K_0$, where the subscript $0$ denotes this particular case, and a factor $\Delta$ is introduced to denote the fraction of the radiation dose invested in this mechanism of DSB, which is proportional to the LET ($\Delta=a\varepsilon$). Likewise, the average number of DSBs produced by two coincident hits $\beta$ comes from a similar procedure. In this case, the DBS probability builds up as the product of two similar expressions, with subscripts 1 and 2 (for each SSB), times the coincidence factor $\xi$, counting for the spatial proximity. A factor $(1-\Delta)$ quantifies the proportion of the dose that leads to DSB through this mechanism \cite{Chadwick}. With these elements, the average number of DSBs per unit of dose, reads for each of the two mechanisms

\begin{align}
	\label{a}	\alpha &= f_0 \Delta \:n_0K_0 \\
	\nonumber \label{b}	\beta &= \xi n_1n_2f_1 f_2 K_1 K_2(1-\Delta)^2,
\end{align}	

\noindent and the total average number of DSBs gets $N_d=\alpha D+\beta D^2$. Now, the survival probability is formulated from the Poisson distribution $P(n)=\frac{(N_d)^ne^{-N_d}}{n!}$, as the probability of not having strand-break events

\begin{equation}
	S(D)=P(0)=e^{-N_d}=e^{-(\alpha D+\beta D^2)}.
	\label{LQ}
\end{equation}

This formulation of Chadwick and Leenhouts establishes the molecular and statistical basis of the very well known LQ-model, shown in equation \ref{LQ}.\\

\subsection{From DNA rupture to cell death}

If we define a critical bond as being located within a target molecule, i.e., $n = n_m$, then a direct comparison between Eqs.~\ref{N5} and \ref{k} reveals that both formulations presented in Eqs.~\ref{beta3} and \ref{a} are mathematically equivalent. Under this assumption, the reaction rates $K_i$ in Eq.~\ref{a} coincide with the rate $K_m$ defined in Eq.~\ref{alpha}, thereby demonstrating the consistency between the probabilistic interpretation of the radiobiological parameters $\alpha$ and $\beta$ and their connection to the physical quantities $\sigma_d, k, \varepsilon, \rho_0, \zeta$, and $dr_i$.

\subsection{Thermal enhancement ratio as a function of hyperthermia}\label{ters}

The radiosensitizing effect of hyperthermia is quantified by the thermal enhancement ratio (TER), defined as the ratio of the radiation dose required to achieve a specific therapeutic effect with radiotherapy alone ($D$) to the dose needed to produce the same effect when RT is combined with HT ($D^*$). Since the addition of HT reduces the required radiation dose ($D^*\le D$). Consequently, the TER, which depends on the HT parameters (treatment time $t$ and temperature $T$), is always greater than or equal to one

\begin{equation}\label{ter}
	\text{\textit{TER}}(T,t)=\frac{D}{D^*}\geq 1.
\end{equation}

By solving for $D=D^*\text{\textit{TER}}$, and substituting it into the LQ-model (Eq.\ref{LQ}) we get

\begin{align*}
	S(D)&=\exp\left\lbrace {-\left[ \alpha(D^*\text{\textit{TER}})+\beta(D^*\text{\textit{TER}})^2\right] }\right\rbrace \\
	&= \exp\left\lbrace {-\left[ \alpha^* D^*+\beta^*(D^*)^2\right]}\right\rbrace .
\end{align*}

\noindent From this expression it can be seen that the new modulated radiobiological parameters are 

\begin{align*}
	\alpha^*(T,t)&=\alpha \text{\textit{TER}}(T,t) \text{   and}\\
	\beta^*(T,t)&=\beta [\text{\textit{TER}}(T,t)]^2,
\end{align*}

\noindent which leaves us with two versions of \textit{TER} that should be equivalent:

\begin{align}
	\label{tera} \text{\textit{TER}}(T,t)&=\frac{\alpha^*(T,t)}{\alpha} \text{   or}\\
	\label{terb} \text{\textit{TER}}(T,t)&=\sqrt{\frac{\beta^*(T,t)}{\beta}}.
\end{align}

For simplicity, we adopt the formulation derived from Eq.~\ref{tera}, where the substitution of Eq.~\ref{alpha3} yields

\begin{equation}\label{ter1}
	\text{\textit{TER}}(T,t)=\frac{(k\, \epsilon\, n_{0m}\, f_0\, \sigma_d\, \rho\, r_i)^*}{k\, \epsilon\, n_{0m}\, f_0\, \sigma_d\, \rho\, r_i},
\end{equation}

\noindent in which the recombination correction has been neglected. This approximation is justified by the fact that the characteristic time scales of the chemical stage are significantly shorter than those required for thermalization, making the contribution of recombination negligible in the context of thermal radiosensitization. Once more, the superscript ($^*$) refers to the parameters that have been modified due to the effect of HT, and some parameters are grouped for convenience. \\
\vspace{0.5cm}

To identify how the thermal enhancement ratio (TER) depends on hyperthermia (HT) parameters --namely temperature ($T$) and treatment duration ($t$)-- we first determined which components of the TER expression are temperature- or time-dependent. Recall that TER is expressed in terms of the following factors: $k\epsilon = dn_i/dl$ (the number of ions generated along the radiation track), $n_{0m}f_0$ (the number of unrepaired target molecules), $\sigma_d$ (the DNA-ion collision cross-section), $\rho$ (the density of the medium, primarily water), and $r_i$ (the diffusion distance of ions after their creation). The functional dependence of each of these factors is defined in section \ref{sensec}.

\subsection{Depletion of DNA repair in cells under HT}\label{Td}

We now analyze the factors that contribute to TER in Equation~\ref{ter1}, starting with the number of vulnerable sites. This term is directly associated with the impairment of DNA repair mechanisms, which is the most widely studied and accepted explanation for thermal radiosensitization. It is important to emphasize that this effect is only observed in cellular systems, where DNA repair pathways are active.\\
\vspace{0.5cm}

An increased number of effective critical sites $n_0f_0$ (bonds that will not be restored by the cellular repair mechanisms) can result from the accumulation of sublethal damage in the cell nucleus, or from an impaired repair processes. Based on the ideas presented in \cite{DeMendoza}, we propose that this number of vulnerable sites increases in a rate-limited manner, because of the denaturation of DNA repair enzymes, or hydrogen bonds at DNA junctions. Accordingly, we propose the relative increase in vulnerable sites to be proportional to the treatment time $t$ and formation rate $\mathcal{K}(T)$:

\begin{equation}
	\frac{(f_0n_0)^*-(f_0n_0)}{(f_0n_0)}=t\mathcal{K}(T).
\end{equation}

Reorganizing, the number of vulnerable sites reads

\begin{equation}\label{misr}
	\frac{(f_0n_0)^*}{f_0n_0}=1+t\mathcal{K}(T),
\end{equation}	

\noindent where $\mathcal{K}(T)=ce^{b(T-T_g)}$ is the  rate of an ``average'' chemical reaction leading to sensitization; with $b$ and $T_g$ being the slope of the temperature-dependent heat capacity, and the melting temperature, respectively. $c\approx1/min$ stands as a frequency coefficient \cite{DeMendoza}. The increased number of critical sites can result in TER values of about 8.0 for C3H cell \textit{in vivo}, depending on the temperature and treatment duration, as shown in reference \cite{Overgaard}.\\
\vspace{0.5cm}

Just considering the contribution of misrepair to TER, previous experimental and theoretical studies have already shown that TER depends exponentially on the treatment temperature. Those results agree with and support the experimental evidence as displayed by Eq.\ref{misr} \cite{DeMendoza}.

\subsection{Cross section of the collision between radiation particles/ions and  DNA} \label{cs}

\begin{figure*}
	\begin{center}
		\includegraphics[width=0.8\textwidth]{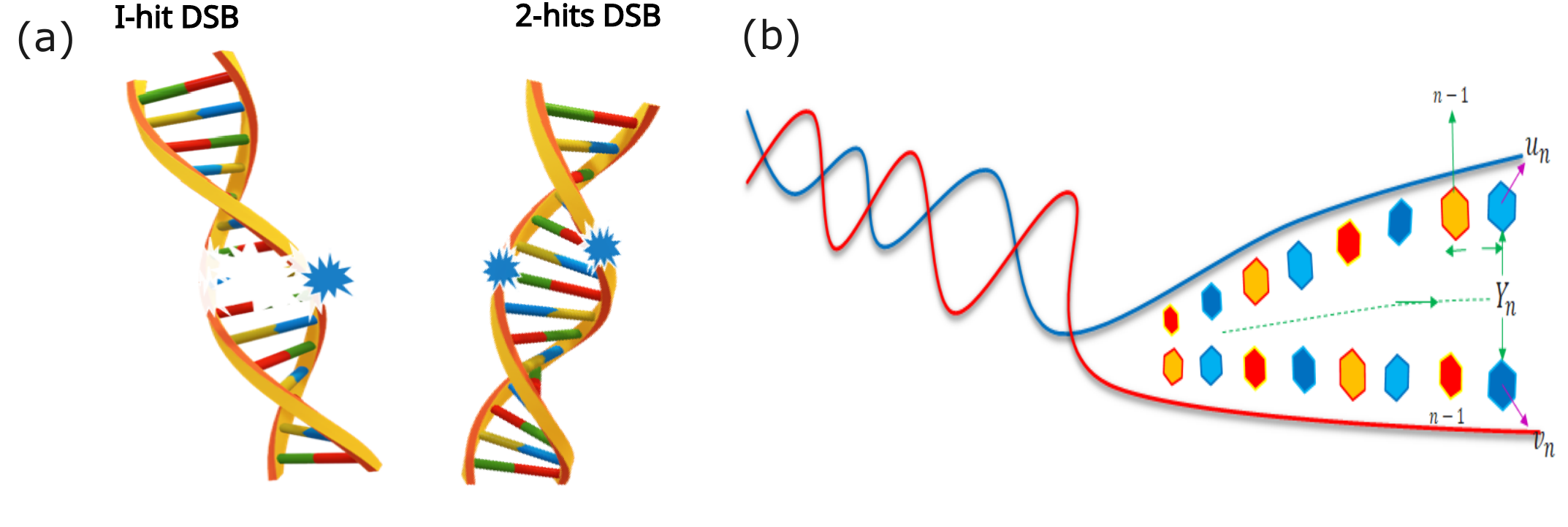}
		\caption{Biophysical mechanisms of DNA double-strand break (DSB) formation and thermal fluctuations. (a) Two distinct pathways can lead to DSBs: a single-hit DSB results from a single radiation particle or ion directly breaking both DNA strands, while a two-hit DSB occurs when two closely spaced single-strand breaks (within ~10 base pairs) lead to a double-strand rupture. (b) Schematic representation of the Peyrard–Bishop model of DNA breathing, illustrating the thermally induced base pair fluctuations that transiently open the DNA double helix. The variables $u_n$ and $v_n$ denote the displacements of bases in a pair, and $Y_n$ the amplitude. Here DNA bases are represented by hexagons. \label{DNAb}}
	\end{center}
\end{figure*}

To model the change in the cross-section of the collision between DNA and ionizing radiation field particles (or intermediate ions), we use the mechanistic-statistical approach of Peyrard and Bishop, which describes the thermal oscillations of DNA \cite{Peyrard1,Dauxois1993}. Peyrard and Bishop's work was motivated by previous Ramman and IR spectroscopy experiments, in which vibrational normal-mode analysis suggested that local denaturation of hydrogen bonds may leads to non-linear breathing dynamics. They proposed a Hamiltonian model, appropriate for the discrete nature of DNA bases, in which the kinetic and potential terms are described as a function of the coordinates $u_n$ and $v_n$ of the nucleotides in each helix respectively (see Figure \ref{DNAb}(b)). \\
\vspace{0.5cm}

Since the aim of this model is to describe the average aperture between the helices, the following transformation is convenient: 

\begin{align}\label{vars}
	Y_n&=\frac{u_n - v_n}{\sqrt{2}},\\
	\nonumber X_n&=\frac{u_n+v_n}{\sqrt{2}}.
\end{align}

\noindent In these coordinates, the Hamiltonian of the molecule reads:

\begin{equation}\label{ham}
	H=\sum_{n=1}^{N}\left\lbrace\frac{{p_n}^{2}}{2m}+\frac{{q_n}^{2}}{2m}+\frac{1}{2}k(X_n-X_{n-1})^2 +\frac{1}{2}k(Y_n-Y_{n-1})^2+D[e^{a\sqrt{2}Y_n}-1]^2\right\rbrace.
\end{equation}

The first 4 terms in the hamiltonian are kinetic and harmonic contributions in X and Y coordinates respectively; the later term accounts for non-linearities, represented by the Morse potential, used to describe the out-of-phase displacements stretching the hydrogen bonds between base-pairs. The partition function is obtained from the Hamiltonian as

\begin{equation}
	Z=\int_{-\infty}^{\infty}\prod_{n=1}^{N}dq_{n} dp_{n} dx_{n}dy_{n}e^{-\boldsymbol{\beta} H(p_{n}...q_{n}...x_{n}...y_{n})}, 
\end{equation}

\noindent where the first four terms of Eq. \ref{ham} contribute with a factor $(2\pi/\boldsymbol{\beta} \omega)^N$, with $\boldsymbol{\beta}=(1/k_BT)$ and $\omega=\sqrt{k/m}$. The anharmonic contribution in the partition function $Z_y$ allows to calculate the average aperture of the DNA molecule like

\begin{equation}\label{yav}
	\left\langle y\right\rangle =\frac{1}{Z_y}\int \prod_N ye^{-\boldsymbol{\beta} f(y_{n}-y_{n-1})}dy_N
\end{equation}

\noindent Here $f(y_{n}-y_{n-1})$ is the anharmonic potential.To this end, the authors applied the transfer integral method, which in the thermodynamic limit ($N \rightarrow \infty$) transforms the computation into the solution of a one-dimensional Schrödinger-type equation \cite{Scalapino,Krumhans,Currie}. The method consists in formulating the transfer integral operator as an eigenvalue equation and solving it \cite{Peyrard1}. See Appendix~\ref{ap1} for details.

\begin{equation}
	\int_{-\infty}^{\infty}dy_{n-1}e^{-\boldsymbol{\beta} f(y_{n}-y_{n-1})}\varphi_i(y_{n-1})=e^{-\boldsymbol{\beta}\varepsilon{i}}\varphi_i (y_{n}).
\end{equation}

After expanding in Taylor series, it is possible to write a Schrödinger-type equation 

\begin{equation}\label{Sch}
	- \frac{1}{2 \beta^2 k} \frac{d^2 \phi(y)}{dy^2} + D (e^{-2\sqrt{2}a y} - 2 e^{-a\sqrt{2} y}) \phi(y) = (\varepsilon - s_0 - D) \phi(y),
\end{equation}
\label{Schr}

\noindent with a constant $s_0=\frac{1}{2\boldsymbol{\beta}}\ln\frac{\boldsymbol{\beta} k}{2\pi} $. 
Expression~\ref{Sch} is a Schrödinger-type equation for a particle in a Morse potential, whose solution is \cite{Peyrard1}

\begin{equation}\label{phii}
	\varphi_n(y)=\mathcal{N}_ne^{-{de^{-\sqrt{2}ay}}}e^{-\sqrt{2}ayR}L_n^{2R}(2de^{-2ay}).
\end{equation}

\noindent See Appendices \ref{ap2} and \ref{ap3} for demonstration. Here $L_n^{2R}$ are the generalized Laguerre polynomials, $d=\frac{\beta \sqrt{kD}}{a}$, and  $R=(d-n-1/2)$. Therefore, the ground-state eigenfunction reads

\begin{equation}\label{phi0}
	\varphi_0=(\sqrt{2}a)^{1/2}\frac{(2d)^{d-\frac{1}{2}}}{\sqrt{\Gamma(2d-1)}}e^{-{de^{-\sqrt{2}ay}}}e^{-\sqrt{2}ay(d-\frac{1}{2})},
\end{equation}

\noindent with its corresponding eigenvalue

\begin{equation*}
	\epsilon_0=\frac{1}{2\boldsymbol{\beta}}\ln\left(\frac{\boldsymbol{\beta} k}{2\pi}\right)+\frac{a}{\boldsymbol{\beta}}\left(\frac{D}{k}\right)^{1/2}-\frac{a^2}{4\boldsymbol{\beta}^2 k}.
\end{equation*}

The integral in Eq.~\ref{yav} can now be evaluated using the result from Eq.~\ref{phi0}. Assuming that the ground state dominates in the thermodynamic limit ($N \rightarrow \infty$), the ground state wave function is employed as a probability amplitude to compute expectation values. This is a standard assumption in statistical mechanics, where the lowest-energy eigenstate typically provides the leading contribution in the large-$N$ limit.

\begin{equation}\label{yav1}
 \left\langle y \right\rangle  =\frac{\sum_i \left\langle \varphi_i(y)|y|\varphi_i(y)\right\rangle}{\sum_i \left\langle \varphi_i(y)|\varphi_i(y)\right\rangle} \simeq \left\langle \varphi_0(y)|y|\varphi_0(y)\right\rangle =\int\varphi_{0}^2(y)ydy
\end{equation}

\begin{figure}[htpb]
	\begin{center}
		\includegraphics[width=0.9\textwidth]{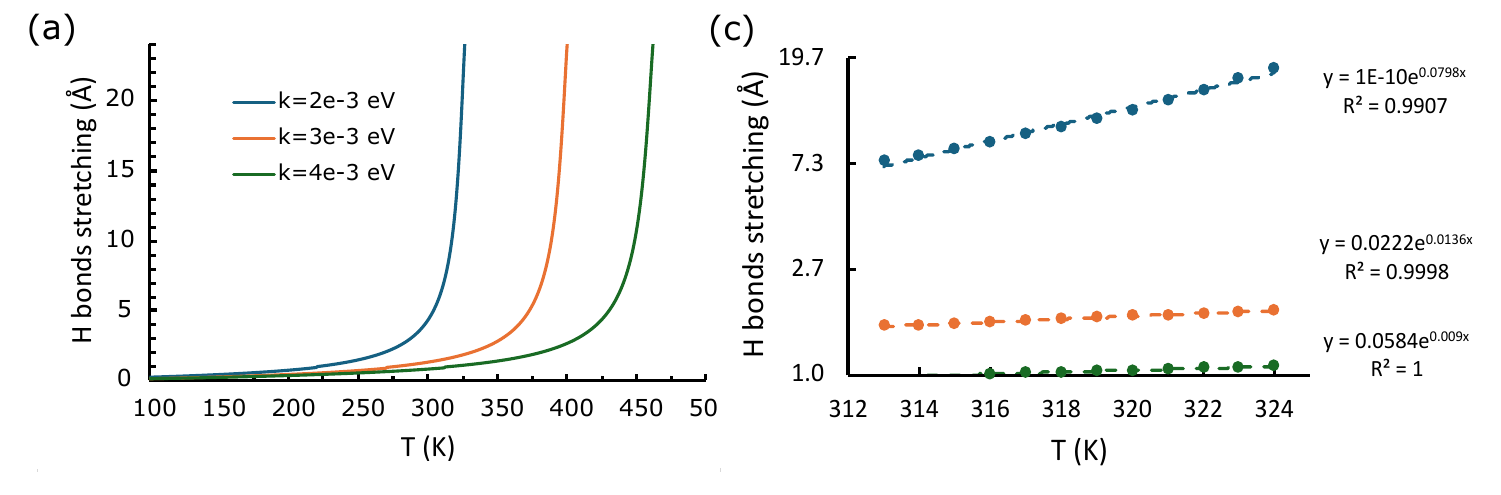}
		\caption{Mean amplitude of DNA thermal fluctuations (DNA-breathing) calculated using the Peyrard–Bishop model \cite{Peyrard1} (Eqs.~\ref{ham} and \ref{yav1}) for three different values of coupling between neighboring bases: $k = 2.0 \times 10^{-3}~\text{eV}/\text{\AA}^2$, $k = 3.0 \times 10^{-3}~\text{eV}/\text{\AA}^2$, and $k = 4.0 \times 10^{-3}~\text{eV}/\text{\AA}^2$. The value of the constant $k$ affects the temperature at which denaturation occurs, and therefore is crucial for the DNA-breathing dynamics. Panel (a) shows the temperature dependence across the full range from 100 to 500~K, while panel (b) focuses on the hyperthermia-relevant range. The logarithmic scale in (b) and the corresponding fitted curves reveal that, within this range, the average strand opening grows approximately exponentially with temperature and can be approximated as $m e^{aT}$. \label{plots}}
	\end{center}
	
\end{figure}

By numerical integration of the Eq. \ref{yav1}, the solid curves presented in Figure~\ref{plots}(a) were obtained. These curves reproduce the curves reported by the authors at temperatures below 400 K such that the model obtained for the average opening of the DNA molecule, under the applied assumptions, is reliable in the HT range between  40$^\circ$C and 50$^\circ$C. To simplify the model, the three curves corresponding to $k = 2 \times 10^{-3}$, $3 \times 10^{-3}$, and $4 \times 10^{-3}$ eV/\AA$^2$ can be approximated by exponential functions of the form $m\,e^{aT}$ (in the HT temperature regime 40-50$^{\circ}$C), with $m$, $a$ being empirical parameters. The values of $m$ are $1 \times 10^{-10}$, $0.0222$, and $0.0584$ eV/\AA$^2$, respectively, while the corresponding values of $a$ are $0.0798$, $0.0136$, and $0.0090$, with $R^2 \approx 1$ in all three cases, as displayed in Fig.\ref{plots}(b).

\subsection{Sensitivity analysis of the TER model}\label{sensec}

To identify which parameters contribute most significantly to the thermal enhancement ratio (TER), we performed a local sensitivity analysis of the main expression of the model (Eq. \ref{ter1}). By expressing the TER as a product of dimensionless ratios, it can be written

\begin{equation*}
	\text{\textit{TER}}(T,t) = \frac{(k\epsilon)^*(T)}{k\epsilon} \cdot \frac{(n_{0m} f_0)^*(T,t)}{n_{0m} f_0} \cdot \frac{\sigma_d^*(T)}{\sigma_d} \cdot \frac{\rho^*(T)}{\rho} \cdot \frac{r_i^*(T,t)}{r_i}.
\end{equation*}

To evaluate the sensitivity of TER to variations in each parameter $x$, we define the normalized (logarithmic) sensitivity index:

\begin{equation}
	S_x = \frac{\partial \log(\text{TER})}{\partial \log(x)} = \frac{x}{\text{TER}} \cdot \frac{\partial\, \text{TER}}{\partial x}.
\end{equation}

\noindent This index quantifies the fractional change in \textit{TER} resulting from a fractional change in the parameter $x$. The advantage of this approach is that it yields dimensionless values that allow direct comparison between parameters with different units and scales. Furthermore, it captures the local proportionality between each parameter and the response function, with $S_x = 1$ indicating direct proportionality, and $S_x = 0$ indicating insensitivity. This type of analysis is particularly useful for identifying the dominant contributors to the thermal enhancement effect under small perturbations of model parameters.\\
\vspace{0.5cm}

We now introduce the functional dependencies of each parameter with respect to temperature $T$ and treatment time $t$, based on physical or biochemical models:

\begin{itemize}
	\item \textit{Medium density} $\rho(T)$ is modeled by the empirical relation $\rho^*(T) = \frac{\rho}{1 + B \Delta T}$, where $B = 2 \times 10^{-4}\;[^\circ\text{C}]^{-1}$ and $\Delta T = T - 37^\circ$C. This relation is independent of treatment time, under the assumption that density remains constant once thermal equilibrium is reached.
	
	\item \textit{Ion diffusion distance} $r_i(T)$ is governed by the Einstein relation for diffusion $r_i^2 = 2 \mathcal{D}(T)t, \quad \text{with} \quad \mathcal{D}(T) = \mu k_B T$. Since HT and RT are applied simultaneously, time $t$ cancels out when computing TER. This leads to the ratio $\frac{r^*_i}{r_i} = \sqrt{\frac{T}{T_0}}$.
	
	\item \textit{Ion generation rate} $k\epsilon(T)$ is modeled using an Arrhenius-type expression $k\epsilon = t A \exp{\left(-\frac{E_a}{k_BT}\right)}$, where $A$ is a pre-exponential factor and $E_a$ the activation energy. Once again, $t$ cancels out in the TER formulation.
	
	\item \textit{DNA collision cross-section} $\sigma_d(T)$ is approximated as an exponential function of temperature, $\sigma_d(T) = m e^{aT}$, based on the Peyrard-Bishop statistical mechanics framework~\cite{Peyrard1}, which describes the increase in DNA breathing amplitude with temperature.
	
	\item \textit{Effective number of vulnerable sites} $(n_{0m}f_0)^*(T)$ is modeled as $1 + t c e^{b(T - T_g)}$, following a formulation proposed in~\cite{DeMendoza} to capture the temperature- and time-dependent denaturation of DNA repair proteins and enzymes. 
	
\end{itemize}

\noindent Hence, keeping just the two main contributors, cross section of damage and critical sites, Eq.\ref{ter1} for the TER function reduces to

\begin{equation}\label{terapp}
	\text{\textit{TER}}(T,t) \simeq \exp[a(T - T_0)] \cdot (1 + t\,c e^{b(T - T_g)}),
\end{equation}

\noindent where $T_0$ is the physiological temperature (37$^\circ$C), $a$ depends on the strength of the Morse potential, and $b$ and $c\approx1/min$ are adjustable parameters described in section \ref{Td} \cite{DeMendoza}. We then derive sensitivity indices with respect to temperature and time:

\begin{align}
	\nonumber	S_T &= \frac{T}{\text{TER}} \cdot \frac{\partial \text{TER}}{\partial T}, \\
	S_t &= \frac{t}{\text{TER}} \cdot \frac{\partial \text{TER}}{\partial t},
\end{align}

\noindent with $\frac{\partial \text{TER}}{\partial T} = e^{a(T - T_0)}\left[a+tce^{b(T - T_g)}(a+b)\right]$, and $\frac{\partial \text{TER}}{\partial t} = ce^{a(T - T_0) + b(T - T_g)}$.\\

\vspace{0.5cm}

These sensitivity indexes can be evaluated numerically and plotted across the therapeutic range of $T$ (40-50$^\circ$C) and $t$ (10-60 min) to identify the dominant mechanisms contributing to TER. This analysis helps to prioritize which biophysical processes (e.g., cross-section growth vs. repair inhibition) are the most relevant in different clinical settings.

\section{Results and discussion}

\subsection{Relative contributions of TER components in the 40--50$^\circ$C range}

In the methods section of this study, we demonstrated that, beyond the impairment of DNA repair mechanisms, additional temperature-dependent biophysical factors may contribute to the synergistic effect between ionizing radiation and hyperthermia. We now implement the functional forms of each relevant variable --namely the temperature or time dependence of medium density $\rho^*$, diffusion distance $r^*_i$, DNA-ion collision cross-section $\sigma^*_d$, ion generation rate $(k\epsilon)^*$, and effective number of vulnerable sites $(n_{0m}f_0)^*$-- and evaluate their relative contributions to the thermal enhancement ratio (TER) within the therapeutic hyperthermia range of 40--50$^\circ$C.\\

\vspace{0.5cm}

Medium density $\rho$ decreases by approximately 0.1\% across this range, indicating a negligible influence on TER. Similarly, the diffusion distance $r_i$ varies by less than 2.1\%, making only a minor contribution. The temperature dependence of ion generation, $k\epsilon$, based on Arrhenius kinetics and activation energies reported in~\cite{Ramos}, results in changes smaller than 0.03\%. While these findings suggest a minimal effect on TER from radiolytic ROS production, it is important to acknowledge that biological systems may experience significant ROS increases under hyperthermia due to non-radiolytic mechanisms such as thermal stress~\cite{Kassis} and the heat-induced expression of endonucleases~\cite{Alberts}, which are not accounted for in the present model. The collision cross-section $\sigma_d$, calculated using the Peyrard-Bishop statistical mechanics framework~\cite{Peyrard1}, increases by approximately 3--5\% under mild hyperthermia, and although insufficient to explain large experimental TER values in cellular systems, its contribution is non-negligible. In contrast, the effective number of vulnerable DNA sites $(n_{0m}f_0)^*$ shows the strongest temperature dependence. Above 41$^\circ$C, the frequency of local base pair openings (DNA breathing) rises exponentially, significantly enhancing the probability of DNA strand breakage by ionizing radiation. This supports the interpretation that DNA structural fluctuations represent a key physical mechanism underlying thermal radiosensitization.\\
\vspace{0.5cm}

By retaining only the two dominant contributors --$\sigma_d$ and $(n_{0m}f_0)^*$-- the TER formulation simplifies to Equation~\ref{terapp}, which preserves the exponential temperature dependence and linear dependence on treatment time. This form is in strong agreement with experimental observations~\cite{Overgaard,Dikomey}, reinforcing the hypothesis that radiosensitization under hyperthermia arises from both impaired DNA repair and enhanced structural vulnerability of DNA due to thermal fluctuations.\\
\vspace{0.5cm}

\begin{figure}
	\includegraphics[width=1.\textwidth]{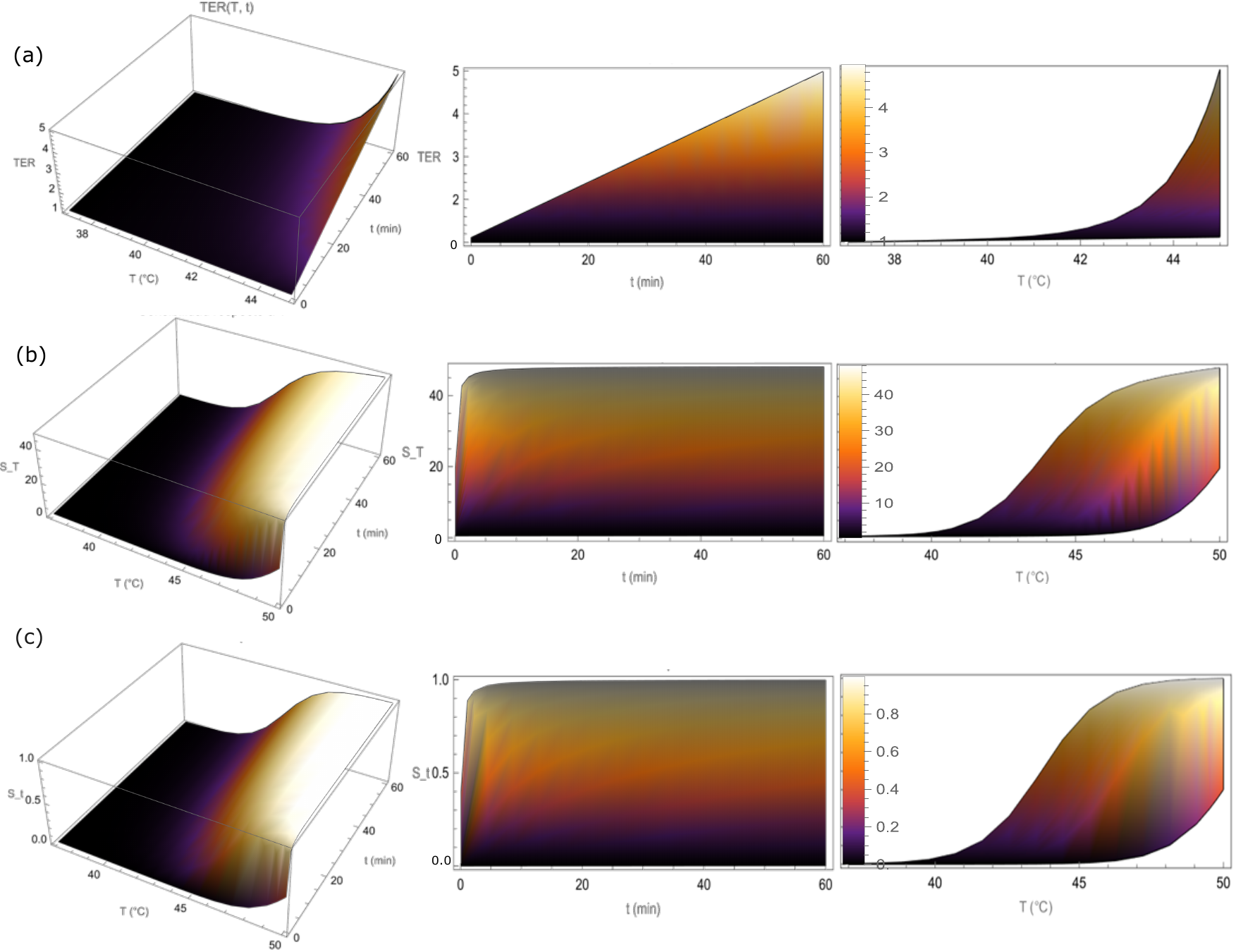}
	\caption{Sensitivity analysis of the thermal enhancement ratio (TER). Panel (a) shows the evolution of TER as a function of temperature and hyperthermia (HT) treatment duration. Multiple views are provided to enhance visualization of the surface topology, highlighting the exponential dependence on temperature and the linear dependence on treatment time. Panels (b) and (c) display the sensitivity indices with respect to temperature ($S_T$) and treatment time ($S_t$), respectively. Again, three perspectives are presented for clarity. Both indices increase with temperature and time, exhibiting similar qualitative trends; however, their magnitudes differ significantly, revealing a much stronger dependence of TER on temperature than on treatment time. \label{sensitivity}}
\end{figure}

Figure~\ref{sensitivity}(a) illustrates the behavior of the thermal enhancement ratio (TER) as a function of hyperthermia temperature $T$ and treatment time $t$. The three panels offer complementary views: a 3D surface plot and two projections along each independent variable. The surface plot shows that TER increases monotonically with both temperature and time. However, the projections clearly reveal different dynamics: while TER grows approximately linearly with treatment time at fixed temperature (center panel), it exhibits an exponential-like increase with temperature, particularly beyond 41$^\circ$C (right panel). This asymmetry highlights that temperature has a stronger influence on the enhancement of radiotherapy efficacy than time, consistent with the sensitivity analysis. The TER remains close to 1 under normothermic conditions or for very short treatment durations, reinforcing the requirement of sustained hyperthermia at elevated temperatures to achieve significant radiosensitization.\\
\vspace{0.5cm}

Figures~\ref{sensitivity}(c,d) summarize the sensitivity analysis of the model with respect to temperature and time. Panel (b) shows the temperature sensitivity index $S_T$, while panel (c) displays the time sensitivity index $S_t$, each from three different viewpoints for clarity. Both indices increase with temperature and time, exhibiting similar qualitative trends but differing significantly in magnitude. Regardless of temperature, both indices rise rapidly and saturate within the first 10--20 minutes of treatment. Sensitivity to both hyperthermia temperature and duration increases sharply at moderate temperatures and tends to plateau above 47$^\circ$C. Although the overall trends are comparable, the magnitudes are markedly different. A comparison of the scales reveals that TER is far more sensitive to changes in temperature than to treatment duration: $S_T$ reaches values approximately 40 times greater than $S_t$, which remains below 1. The cross-sectional views in (b) further emphasize that temperature has a dominant influence on the model’s response, particularly beyond $T = 41^\circ$C. These results support the interpretation that thermal enhancement is primarily driven by temperature-dependent biophysical mechanisms, such as the inhibition of DNA repair, with treatment time playing a secondary, though still relevant, role.\\

\vspace{0.5cm}
The greater sensitivity of TER to changes in temperature compared to treatment time can be attributed to the exponential nature of the two main biophysical processes driving thermal radiosensitization: the denaturation of DNA repair proteins and the increased susceptibility of DNA due to enhanced thermal fluctuations. As temperature increases, protein denaturation accelerates sharply, thereby compromising the cell's ability to repair radiation-induced DNA damage. At the same time, higher thermal energy amplifies the vibrational motion of DNA strands, increasing their vulnerability to double-strand breaks. As discussed in the Methods section leading to Equation~\ref{terapp}, both processes are thermally activated and exhibit exponential dependence on temperature, resulting in a strong amplification of the radiosensitizing effect. In contrast, time contributes linearly to the accumulation of unrepaired damage, provided that the impairment of repair mechanisms has already been established. This fundamental difference in the underlying mathematical behavior of the mechanisms explains the predominant influence of temperature on TER.

\subsection{TER in isolated plasmids}

It is important to note that the effects of heat on DNA repair proteins can only be observed in cell-based experiments, where the corresponding enzymatic machinery is active. In such systems, the thermal enhancement ratio (TER) has been widely reported. However, to isolate and quantify the contribution of other temperature-dependent factors --independent of repair inhibition-- cell-free assays using purified DNA have been employed. These experiments allow for the direct measurement of DNA damage as a function of temperature. To our knowledge, only one study --by Tomita et al.~(1995)~\cite{Tomita}-- has reported both single- and double-strand breaks in isolated plasmid DNA exposed to $\gamma$-radiation at temperatures ranging from -20$\,^\circ$C to 42$\,^\circ$C. In their work, dilute aqueous solutions of plasmid DNA (29.75\,$\mu$g\,cm$^{-3}$ DNA, 1\,mmol\,dm$^{-3}$ Tris, 5\,mmol\,dm$^{-3}$ NaCl, and 0.1\,mmol\,dm$^{-3}$ EDTA) were irradiated with $^{60}$Co $\gamma$-rays, and the samples were subsequently analyzed by agarose gel electrophoresis. This technique allowed them to quantify the relative proportions of supercoiled (intact), nicked (single-strand breaks), and linearized (double-strand breaks) DNA forms. Their results revealed a clear temperature-dependent increase in radiation-induced strand breaks, even in the absence of cellular repair processes.\\
\vspace{0.4cm}

To investigate the quantitative contribution of thermally induced DNA destabilization to radiosensitization, we compared both experimental and analytical estimates of the thermal enhancement ratio (TER), defined as the relative increase in DNA damage yield at elevated temperatures. Experimental TER values were derived from the data reported by Tomita et al., calculated as the ratio of single-strand break (SSB) induction efficiencies at a given temperature relative to a reference value $T_0$.\\

\begin{figure}[ht]
	\centering
	\includegraphics[width=0.75\textwidth]{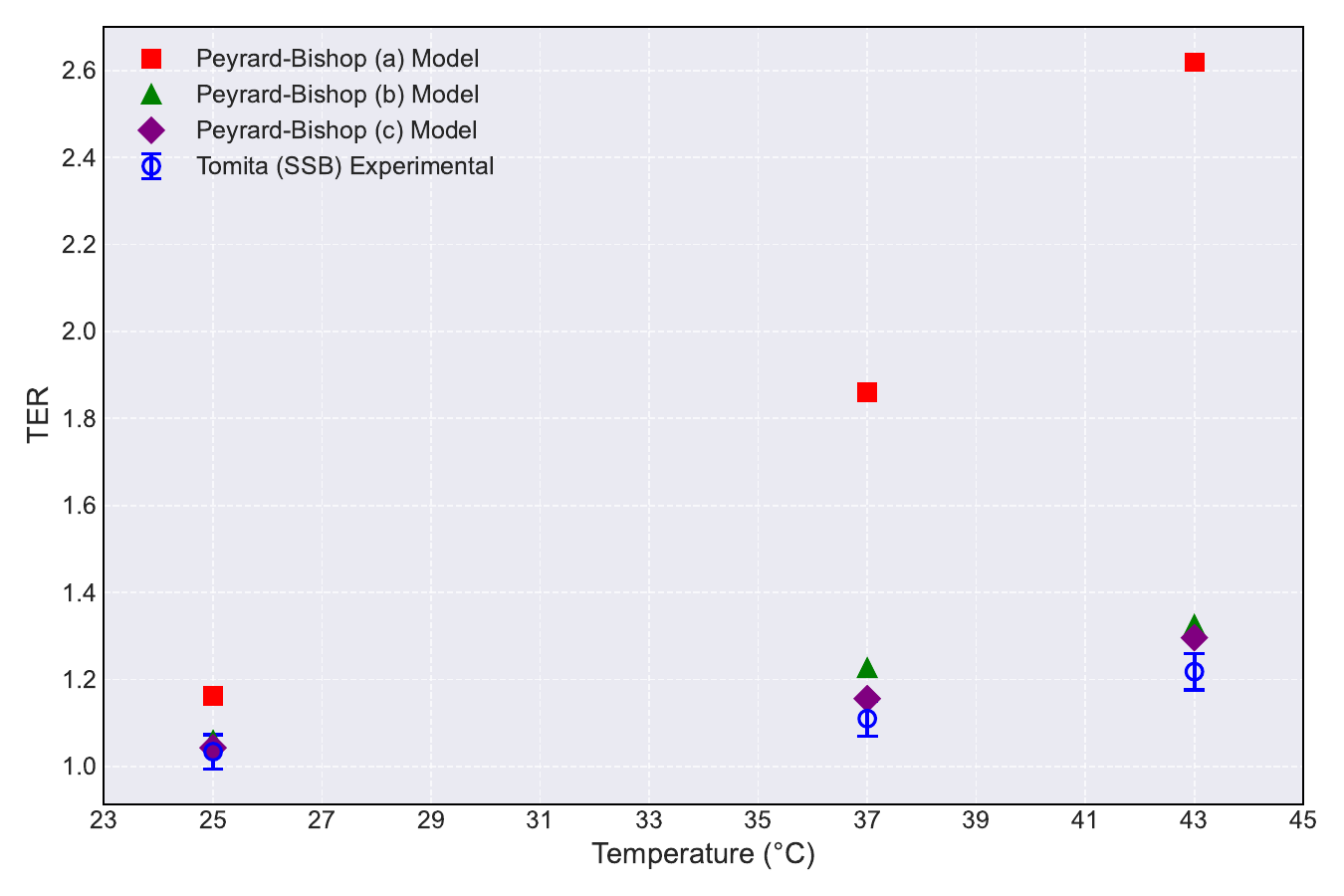}
	\caption{Temperature Enhancement Ratio (TER) as a function of temperature for different values of the stiffness constant \(k\) in the Peyrard-Bishop model, labeled as (a), (b), and (c), respectively. TER values were normalized with respect to their corresponding values at 20°C. Model predictions are compared against experimental TER values obtained from Tomita et al.~\cite{Tomita}, based on single-strand break (SSB) data. The error bars for the experimental data represent the standard deviations estimated from the original study.}
	\label{TERPBTomita}
\end{figure}

\begin{table}[htbp]
  \centering
  \caption{Comparison of TER Values and Tomita/P-B Model Ratios at Different Reference Temperatures $T_0$.}
    \label{TERtab}
    \begin{tabular}{l cc c cc}
        \toprule
        \multirow{2}{*}{\textbf{Model}} & \multicolumn{2}{c}{\textbf{$T_0 = 25 \, ^\circ\text{C}$}} & \phantom{abc} & \multicolumn{2}{c}{\textbf{$T_0 = 37 \, ^\circ\text{C}$}} \\
        \cmidrule{2-3} \cmidrule{5-6}
        & \textbf{TER Value} & \textbf{Ratio Tomita/P-B} & & \textbf{TER Value} & \textbf{Ratio Tomita/P-B} \\
        \midrule
        Tomita(SSB) & $1.178 \pm 0.046$ & N/A & & $1.097 \pm 0.041$ & N/A \\
        P-B(a)      & $2.253 $    & $0.523 \pm 0.020$ & & $1.407 $    & $0.779 \pm 0.029$ \\
        P-B(b)      & $1.250 $    & $0.942 \pm 0.037$ & & $1.081 $    & $1.014 \pm 0.038$ \\
        P-B(c)      & $1.243 $    & $0.948 \pm 0.037$ & & $1.120 $    & $0.978 \pm 0.036$ \\       
        \bottomrule
    \end{tabular}
\end{table}

\vspace{0.4cm}

In our model, TER was evaluated analytically from Eq.~\ref{ter1}, which considers multiple mechanistic contributors to radiosensitization. Among these, DNA breathing --quantified using the Peyrard–Bishop (P-B) model-- emerges as the second most influential factor after misrepair. Its contribution was estimated as the ratio of mean thermal fluctuation amplitudes at the relevant temperatures. Figure~\ref{TERPBTomita} shows the temperature dependence of the normalized TER predicted by the P--B model for three values of the coupling constant between neighboring bases, $k$: (a)~$2.0 \times 10^{-3}~\text{eV}/\text{\AA}^2$, (b)~$3.0 \times 10^{-3}~\text{eV}/\text{\AA}^2$, and (c)~$4.0 \times 10^{-3}~\text{eV}/\text{\AA}^2$. All curves are normalized with respect to TER at 20~$^\circ$C to highlight the relative trends without overcrowding the figure. The choice of $k$ significantly affects the predicted magnitude of DNA breathing and, consequently, the radiosensitization response. Model~(a), corresponding to the weakest coupling, overestimates TER at higher temperatures, suggesting unrealistically large base-pair fluctuations. In contrast, model~(c), with the strongest coupling, yields predictions more consistent with the experimental trend, indicating that moderate coupling values offer a more realistic thermal response.\\
\vspace{0.4cm}

To extend the evaluation of model performance to other temperature conditions (albeit with fewer data points), Table~\ref{TERtab} presents absolute TER values at 43~$^\circ$C, calculated relative to both 25~$^\circ$C and 37~$^\circ$C. The analytical TER values were calculated again isolating the contribution from DNA breathing. At $T_0 = 37^\circ$C, the relative deviations for models (a), (b), and (c) are 28.3\%, 1.5\%, and 2.2\%, respectively. For $T_0 = 25^\circ$C, the discrepancies increase to 91.2\%, 6.2\%, and 5.5\%. The “Ratio Tomita/P-B” columns in Table~\ref{TERtab} quantify these differences explicitly, confirming that model accuracy improves with increasing $k$.\\ 
\vspace{0.4cm}

These findings support the hypothesis that thermally enhanced DNA breathing, as described by the P--B model, plays a measurable and quantifiable role in radiosensitization. Although it alone does not account for the full magnitude of TER observed experimentally, it constitutes a robust second-order effect, complementing misrepair mechanisms in shaping the overall response.\\
\vspace{0.4cm}

Our model offers a mechanistic insight into the molecular basis of radiosensitization under concurrent hyperthermia and radiotherapy. It assumes uniform tissue heating and simplifies complex biological responses using effective temperature- and time-dependent parameters. Although it does not include spatial heterogeneities, sequential treatment effects, or systemic responses such as immune modulation or vascular changes, it serves as a valuable foundation for understanding thermal enhancement at the molecular level. Future work may expand the model toward more comprehensive, clinically relevant scenarios.\\

\vspace{0.4cm}
To further validate the predictions of our model, future experimental efforts should focus on the concomitant application of hyperthermia and ionizing radiation to isolated DNA or plasmids, varying both radiation dose/LET and HT temperature/time. Such experiments would enable a more direct quantification of thermal enhancement under controlled conditions in the absence of cellular DNA repair mechanisms. Techniques such as single-molecule Förster Resonance Energy Transfer (smFRET) have proven effective for probing local DNA breathing dynamics at base-pair resolution~\cite{AltanBonnet2003,Jose2012}, while UV absorbance methods exploiting the hyperchromic effect have been widely used to quantify DNA strand separation and melting transitions~\cite{DAbramo}. These approaches could be combined with strand break assessments—e.g., via agarose gel electrophoresis—to directly correlate thermally induced DNA fluctuations with radiosensitization outcomes. Such measurements would provide valuable insight into the temperature-dependent amplification of DNA vulnerability proposed in this study. Given the current scarcity of experimental data at the molecular level, we plan in the near future to conduct Monte Carlo simulations in which the temperature dependent effects of hyperthermia are implemented in reaction rates, diffusion coefficients, medium density, DNA repair, and DNA-ion collision cross sections. These simulations aim to quantitatively explore the impact of thermal parameters on radiosensitization and guide the design of future \textit{in vitro} and \textit{in vivo} validation studies.

\section{Conclusions}

In this work, we developed a simplified mathematical model to describe the thermal enhancement ratio (TER) of radiotherapy as a function of temperature and treatment time. The model captures the two principal mechanisms contributing to radiosensitization under hyperthermia: (1) the exponential increase in DNA vulnerability due to enhanced thermal oscillations (DNA breathing), and (2) the rapid denaturation of DNA repair proteins, both of which are thermally activated processes. As a result, TER exhibits an exponential dependence on temperature and a linear dependence on time, consistent with experimental observations in the literature. Importantly, the increased susceptibility of DNA to ionizing radiation through amplified thermal oscillations occurs only during the concomitant application of hyperthermia and radiation. This phenomenon may partly explain the pronounced peak in TER observed experimentally when HT and RT are administered simultaneously.\\
\vspace{0.4cm}

Sensitivity analysis revealed that TER is significantly more responsive to changes in temperature than to treatment time. This asymmetry stems from the exponential nature of the underlying thermal effects on biomolecular stability, while the contribution of time arises primarily through the linear accumulation of unrepaired DNA damage once repair mechanisms are compromised. The model also shows that TER saturates with time after approximately 10-20 minutes of hyperthermia, particularly at temperatures above $43^\circ$C, suggesting a plateau in radiosensitization once repair inhibition is maximal. Overall, the results support the interpretation that temperature is the dominant driver of thermal radiosensitization, reinforcing the importance of precise thermal control during clinical hyperthermia. Moreover, the compact analytical form of the model provides a useful tool for interpreting experimental results and for guiding the design of optimized combined hyperthermia-radiotherapy protocols. This model may also serve as a foundational component for more complex simulations at the tissue or tumor scale, offering a mechanistic basis for integrating molecular effects into multiscale treatment planning frameworks.

\funding{This research did not receive external funding.}

\roles{All authors contributed equally to the conceptualization of the study, validation of the results, and interpretation of the findings. J.L.R. and A.M.D.M. were responsible for the development and implementation of the analytical and numerical calculations. All authors jointly participated in the investigation and in the preparation of the manuscript, including the creation of figures and tables. All authors have reviewed and approved the final version of the manuscript.}

\data{Data Availability: No new data were generated for this study. All datasets used for model calibration, illustration, or validation were obtained from previously published sources, which are cited accordingly within the manuscript. These data are publicly available in the corresponding publications. Further information is available from the corresponding author (Adriana De Mendoza, a.demendoza@javeriana.edu.co)  upon reasonable request.}

\suppdata{Not applicable.}\\

\vspace{0.4cm}
\noindent \textbf{AI assistance}\\
\noindent The authors acknowledge the use of OpenAI’s ChatGPT (GPT-4, June 2024 version) for assistance with language refinement, technical editing, LaTeX formatting, and bibliography management. All AI-generated content was reviewed and edited by the authors, who take full responsibility for the final manuscript.\\
\vspace{2cm}

\appendix

\textbf{{\Large Appendices}}\\

\section{Transfer integral method}\label{ap1}
\setcounter{section}{1}
\renewcommand{\thesection}{\Alph{section}}

	In statistical mechanics, we are often interested in computing the canonical partition function for a one-dimensional chain of $N$ coupled degrees of freedom, such as the inter-strand displacements $y_1, y_2, \dots, y_N$ in a DNA denaturation model. When the Hamiltonian includes nearest-neighbor interactions, it takes the general form:
	
\begin{equation}
	H(\{y_n\}) = \sum_{n=1}^{N} \left[ V(y_n) + W(y_n, y_{n-1}) \right],
\end{equation}	
	
\noindent where $V(y_n)$ is an on-site potential (e.g., Morse potential) and $W(y_n, y_{n-1}) = \frac{k}{2}(y_n - y_{n-1})^2$ represents harmonic coupling between neighboring sites. The canonical partition function is then:
	
\begin{equation}
	Z_y = \int dy_1 \cdots dy_N \, \exp\left[ -\beta H(\{y_n\}) \right].
\end{equation}	
	
	Because the interaction term $W(y_n, y_{n-1})$ couples neighboring variables, the integrand cannot be factorized into a product of independent terms. Instead, we rewrite it using a transfer kernel:
	
\begin{equation}
	K(y_n, y_{n-1}) = \exp\left[ -\beta \left( \frac{k}{2}(y_n - y_{n-1})^2 + \frac{1}{2} \left( V(y_n) + V(y_{n-1}) \right) \right) \right],
\end{equation}	
	
\noindent which allows the partition function to be expressed as:
	
\begin{equation}
	Z = \int dy_1 \cdots dy_N \, K(y_1, y_2) K(y_2, y_3) \cdots K(y_N, y_1).
\end{equation}	
	
\noindent Under the assumption of periodic boundary conditions ($y_{N+1} = y_1$), the resulting expression is mathematically equivalent to the trace of the $N$-th power of the integral operator $K$:
	
\begin{equation}
	Z = \mathrm{Tr}(K^N),
\end{equation}	
	
%
	
\noindent This formulation is analogous to linear algebra, where the trace of a matrix power is the sum of its eigenvalues raised to that power:
	
\begin{equation}
	\mathrm{Tr}(A^N) = \sum_i \lambda_i^N.
\end{equation}	
	
By analogy, we interpret $K$ as an operator with a discrete spectrum of eigenvalues $\lambda_0, \lambda_1, \dots$, and in the thermodynamic limit ($N \to \infty$), the partition function is dominated by the largest eigenvalue $\lambda_0$:
	
\begin{equation}
	Z \sim \lambda_0^N.
\end{equation}	
	
Thus, the reason the partition function can be expressed as $\mathrm{Tr}(K^N)$ is that the statistical weight of the entire system is governed by repeated applications of the local transfer operator $K$, and the cyclical nature imposed by the boundary conditions leads naturally to the trace structure.

\section{Derivation of the Schr\"odinger-Type Equation from the Transfer Integral Formalism}\label{ap2}
\setcounter{section}{2}
\renewcommand{\thesection}{\Alph{section}}
\addcontentsline{toc}{section}{Appendix B. Derivation of the Schr\"odinger-Type Equation}

In the main text, the partition function for the nonlinear DNA denaturation model is expressed using a transfer integral approach. This leads to the eigenvalue equation for the transfer operator $K(y, y')$:

\begin{equation}
	\int_{-\infty}^{\infty} dy'\; \exp\left[ -\frac{1}{2} \beta k (y - y')^2 - \frac{1}{2} \beta (V(y) + V(y')) \right] \phi(y') = e^{-\beta \varepsilon} \phi(y).
	\label{eq:transfer_integral}
\end{equation}

To analyze this equation in the thermodynamic limit ($N \to \infty$), we approximate the integral kernel as a Gaussian operator. First, we factor out the term that depends only on $y$:

\begin{equation}
	K(y, y') = e^{-\frac{1}{2} \beta V(y)} \cdot e^{-\frac{1}{2} \beta k (y - y')^2} \cdot e^{-\frac{1}{2} \beta V(y')}.
\end{equation}

We now consider the normalized Gaussian convolution operator \cite{Simon}:

\begin{equation}
	\int_{-\infty}^{\infty} dy'\; e^{- \frac{\beta k}{2} (y - y')^2} \psi(y') = \sqrt{\frac{2\pi}{\beta k}} \left( e^{\frac{1}{2 \beta k} \frac{d^2}{dy^2}} \psi \right)(y),
\end{equation}

\noindent which is valid if $\psi(y)$ is smooth. Applying this to the full equation, we obtain:

\begin{equation}\label{gauss}
	e^{-\frac{1}{2} \beta V(y)} \left( e^{\frac{1}{2 \beta k} \frac{d^2}{dy^2}} \left[ e^{-\frac{1}{2} \beta V(y)} \phi(y) \right] \right) = \sqrt{\frac{\beta k}{2\pi}} e^{-\beta \varepsilon} \phi(y).
\end{equation}

Now, we expand the exponential operator to second order, valid in the limit $\beta k \gg 1$ (strong coupling or low temperature):

\begin{equation*}
	e^{\frac{1}{2 \beta k} \frac{d^2}{dy^2}} \approx 1 + \frac{1}{2 \beta k} \frac{d^2}{dy^2},
\end{equation*}

\noindent and apply this approximation to Equation \ref{gauss}:

\begin{equation}
	e^{-\frac{1}{2} \beta V(y)} \left( 1 + \frac{1}{2 \beta k} \frac{d^2}{dy^2} 
	\right) \left[ e^{-\frac{1}{2} \beta V(y)} \phi(y) \right] \approx \sqrt{\frac{\beta k}{2\pi}} e^{-\beta \varepsilon} \phi(y).
\end{equation}

Next, we multiply both sides by $e^{\frac{1}{2} \beta V(y)}$ and evaluate the derivatives:

\begin{align}
	\left( 1 + \frac{1}{2 \beta k} \left[ -\frac{1}{2} \beta V''(y) + \frac{1}{4} \beta^2 (V'(y))^2 - \beta V'(y) \frac{d}{dy} + \frac{d^2}{dy^2} \right] \right) \phi(y) \approx \sqrt{\frac{\beta k}{2\pi}} e^{-\beta \varepsilon + \frac{1}{2} \beta V(y)} \phi(y).
\end{align}

Among the terms in the differential operator, the terms involving $V'(y)$ and $V''(y)$ behave as perturbative corrections to the potential, and are therefore neglected under the leading-order semi-classical approximation. Taking the logarithm of both sides, we obtain:

\begin{equation}
\ln \left( \frac{\left(1 + \frac{1}{2 \beta k} \hat{O} \right) \phi(y)}{\phi(y)} \right)  = -\beta \varepsilon + \frac{\beta}{2} V(y) + \ln \left( \sqrt{\frac{\beta k}{2\pi}} \right),
\end{equation}

\noindent where $\hat{O}$ is the differential operator in parentheses. Expanding the left-hand side and dividing by $\phi(y)$:

\begin{equation}
	\frac{1}{2 \beta k} \frac{\hat{O} \phi(y)}{\phi(y)} \approx - \beta \varepsilon + \frac{\beta}{2} V(y) + \ln \left( \sqrt{\frac{\beta k}{2\pi}} \right).
\end{equation}

Dividing by $\beta$ and rearranging terms, we obtain the effective Schr\"odinger-type equation:

\begin{equation}
	-\frac{1}{2 \beta^2 k} \frac{d^2 \phi(y)}{dy^2} + \frac{1}{2} V(y) \phi(y) = (\varepsilon - s_0) \phi(y),
\end{equation}

\noindent where $s_0 = \frac{1}{2 \beta} \ln\left( \frac{\beta k}{2\pi} \right)$. Finally, substituting the Morse potential

\begin{equation}\label{EqSch}
	V(y) = D (e^{-a \sqrt{2} y} - 1)^2 = D (e^{-2\sqrt{2}a y} - 2 e^{-a\sqrt{2} y} + 1),
\end{equation}

\noindent the Schr\"odinger type equiation takes the form of Eq. \ref{Sch} 

\begin{equation*}
	- \frac{1}{2 \beta^2 k} \frac{d^2 \phi(y)}{dy^2} + D (e^{-2\sqrt{2}a y} - 2 e^{-a\sqrt{2} y}) \phi(y) = (\varepsilon - s_0 - D) \phi(y).
\end{equation*}

\section{Solution of the Effective Schrödinger-Type Equation}\label{ap3}
\setcounter{section}{2}
\renewcommand{\thesection}{\Alph{section}}
\addcontentsline{toc}{section}{Appendix C. Solution of the Effective Schrödinger-Type Equation}

We consider the effective Schrödinger-type equation \ref{EqSch}, which describes the thermal fluctuations of the base-pair stretching coordinate $y$ in the Peyrard-Bishop model under the semiclassical approximation. To solve it, we perform the change of variable:

\begin{equation}\label{z}
	z = e^{-\sqrt{2}a y},
\end{equation}

\noindent so that derivatives transform as $	\frac{dz}{dy} = -\sqrt{2}a z$, and $\frac{d^2z}{dy^2} = 2a^2 z$. Substituting into Eq.~\ref{EqSch}, we obtain a second-order differential equation in the variable \( z \), which takes the general form:

\begin{equation}\label{laguerre}
	z^2 \frac{d^2 \varphi}{dz^2} + z \frac{d \varphi}{dz} + d^2\left( -\lambda + z - \frac{z^2}{2} \right) \varphi = 0.
\end{equation}

\noindent Here $\lambda, d$ are constants depending on the physical parameters $D, a, \beta, k$

\begin{equation*}
	d=\frac{\beta}{a}\sqrt{kD},\text{\textcolor{white}{------}}
	\lambda = \frac{1}{D}\left( \varepsilon - s_0 - D/2 \right).
\end{equation*}

Equation \ref{laguerre} admits as general solution a linear combination of special functions:

\begin{align}
	\varphi(z) =\; 
	e^{ d \left( -\frac{z}{\sqrt{2}} + \sqrt{\lambda} \log z \right) } \Bigg\{ &
	C_1\,
	\mathrm{U}\left(
	\frac{1}{2} \left( 1 - \sqrt{2} d + 2 d \sqrt{\lambda} \right),\;
	1 + 2 d \sqrt{\lambda},\;
	\sqrt{2} d z
	\right)
	\nonumber \\
	&+ C_2\,
	L^{(2 d \sqrt{\lambda})}_{ \frac{1}{2} (-1 + \sqrt{2} d - 2 d \sqrt{\lambda}) }
	\left( \sqrt{2} d z \right)
	\Bigg\},
\end{align}

\noindent where $U$ is the confluent hypergeometric function of the second kind, and $L_n^R$ is the generalized Laguerre polynomial. To obtain a physically acceptable solution, we impose regularity at $z = 0$ (corresponding to $y \to \infty$), and normalizability over the domain of $y$. Since the function $U$ diverges as $z \to 0$, we discard the term with coefficient $c_1$ and retain only the Laguerre part. Thus, the physical solution is

\begin{equation}\label{phiz}
	\varphi_n(z) = \mathcal{N}_n \, e^{ d \left( -\frac{z}{\sqrt{2}} + \sqrt{\lambda} \log z \right) } L^{(2 d \sqrt{\lambda})}_{ \frac{1}{2} (-1 + \sqrt{2} d - 2 d \sqrt{\lambda})}\left( \sqrt{2} d z \right).
\end{equation}

\noindent The normalization constant $\mathcal{N}_n$ is determined from $\int_0^\infty \varphi_n^2(y)\, dy = 1$. To get Equation~\ref{phii}, we reverse the change of variable (Eq.~\ref{z}), such that Eq.~\ref{phiz} becomes

\begin{equation}
	\varphi_n(y) = \mathcal{N}_n\, e^{-\frac{d}{\sqrt{2}} e^{-\sqrt{2} a y}} \, e^{ - \sqrt{2} a y R } \, L_n^{(2R)}\left( \sqrt{2} d e^{ -\sqrt{2} a y } \right),
\end{equation}

\noindent under the quantization condition $d\sqrt{\lambda} = R = d/\sqrt{2}-n-1/2$. If $d \rightarrow \sqrt{2}d$, Equation~\ref{phii} is recovered.

\providecommand{\newblock}{}

\bibliographystyle{iopart-num}

\begin{thebibliography}{10}
	\expandafter\ifx\csname url\endcsname\relax
	\def\url#1{{\tt #1}}\fi
	\expandafter\ifx\csname urlprefix\endcsname\relax\def\urlprefix{URL }\fi
	\providecommand{\eprint}[2][]{\url{#2}}
	
	\bibitem{Overgaard}
	Overgaard J 1980 {\em International Journal of Radiation
		Oncology*Biology*Physics\/} {\bf 6} 1507--1517 ISSN 0360-3016
	\urlprefix\url{http://www.sciencedirect.com/science/article/pii/0360301680900085}
	
	\bibitem{Horsman}
	Horsman M and Overgaard J 2007 {\em Clinical Oncology\/} {\bf 19} 418--426 ISSN
	0936-6555 importance of Radiobiology to Cancer Therapy: Current Practice and
	Future Perspectives
	\urlprefix\url{http://www.sciencedirect.com/science/article/pii/S0936655507005870}
	
	\bibitem{Overgaard2013}
	Overgaard J 2013 {\em Radiotherapy and Oncology\/} {\bf 109} 185--187
	
	\bibitem{Mei}
	Mei X, ten Cate R, van Leeuwen C~M, Rodermond H~M, de~Leeuw L, Dimitrakopoulou
	D, Stalpers L~J~A, Crezee J, Kok H~P, Franken N~A~P and Oei A~L 2020 {\em
		Cancers\/} {\bf 12} ISSN 2072-6694
	\urlprefix\url{https://www.mdpi.com/2072-6694/12/3/582}
	
	\bibitem{Elming}
	Elming P~B, S{\o}rensen B~S, Oei A~L, Franken N~A~P, Crezee J, Overgaard J and
	Horsman M~R 2019 {\em Cancers\/} {\bf 11} ISSN 2072-6694
	\urlprefix\url{https://europepmc.org/articles/PMC6356970}
	
	\bibitem{van_Leeuwen2017}
	van Leeuwen C~M, Crezee J, Oei A~L, Franken N~A~P, Stalpers L~J~A, Bel A and
	Kok H~P 2018 {\em International Journal of Hyperthermia\/} {\bf 34} 901--909
	\urlprefix\url{https://doi.org/10.1080/02656736.2018.1468930}
	
	\bibitem{van_Leeuwen20172}
	van Leeuwen C~M, Oei A~L, Chin K~W~T~K, Crezee J, Bel A, Westermann A~M, Buist
	M~R, Franken N~A~P, Stalpers L~J~A and Kok H~P 2017 {\em Radiation
	  Oncology\/} {\bf 12} 1--8
        \urlprefix\url{https://doi.org/10.1186/s13014-017-0813-0}
	
	\bibitem{Lindegaard3}
	Lindegaard J~C 1992 {\em International Journal of Hyperthermia\/} {\bf 8}
	561--586 \urlprefix\url{https://doi.org/10.3109/02656739209037994}
	
	\bibitem{Oei}
	Oei A~L, Vriend L~E~M, Crezee J, Franken N~A~P and Krawczyk P~M 2015 {\em
		Radiation Oncology\/} {\bf 165}
	
	\bibitem{Lepock}
	Lepock J~R 2004 {\em International Journal of Hyperthermia\/} {\bf 20} 115--130
	\urlprefix\url{https://doi.org/10.1080/02656730310001637334}
	
	\bibitem{Lepock1}
	Lepock J~R 2005 {\em International Journal of Hyperthermia\/} {\bf 21} 681--687
	\urlprefix\url{https://doi.org/10.1080/02656730500307298}
	
	\bibitem{Lepock2}
	Lepock J, Frey H and Ritchie K 1993 {\em Journal of Cell Biology\/} {\bf 122}
	1267--1276 ISSN 0021-9525
	\urlprefix\url{https://rupress.org/jcb/article-pdf/122/6/1267/385325/1267.pdf}
	
	\bibitem{Roti}
	Roti-Roti J~L 2008 {\em International Journal of Hyperthermia\/} {\bf 24} 3--15
	\urlprefix\url{https://doi.org/10.1080/02656730701769841}
	
	\bibitem{Hildebrandt}
	Hildebrandt B, Wust P, Ahlers O, Dieing A, Sreenivasa G, Kerner T, Felix R and
	Riess H 2002 {\em Critical Reviews in Oncology/Hematology\/} {\bf 43} 33--56
	ISSN 1040-8428
	\urlprefix\url{http://www.sciencedirect.com/science/article/pii/S1040842801001792}
	
	\bibitem{DeMendoza}
	De~Mendoza A~M, Michl\'ikov\'a S, Berger J, Karschau J, Kunz-Schughart L~A and
	McLeod D~D 2021 {\em Scientific Reports\/} {\bf 11} 1 -- 14
	\urlprefix\url{https://doi.org/10.1038/s41598-021-84620-z}
	
	\bibitem{DeMendoza1}
	De~Mendoza A~M, Michlíková S, Castro P~S, Muñoz A~G, Eckhardt L, Lange S and
	Kunz‑Schughart L~A 2025 {\em Physics in Medicine \& Biology\/} {\bf 70}
	025022
	\urlprefix\url{https://iopscience.iop.org/article/10.1088/1361-6560/ada680}
	
	\bibitem{Jungbauer}
	Jungbauer A and Kaar W 2007 {\em Journal of Biotechnology\/} {\bf 128} 587--596
	ISSN 0168-1656
	\urlprefix\url{https://www.sciencedirect.com/science/article/pii/S0168165606010273}
	
	\bibitem{Parsell}
	Parsell D~A, Kowal A~S, Singer M~A and Lindquist S 1994 {\em Nature\/} {\bf
		372} 475--478 ISSN 0168-1656 \urlprefix\url{https://doi.org/10.1038/372475a0}
	
	\bibitem{Seckler}
	Seckler R and Jaenicke R 1992 {\em The FASEB Journal\/} {\bf 6} 2545--2552
	\urlprefix\url{https://faseb.onlinelibrary.wiley.com/doi/abs/10.1096/fasebj.6.8.1592207}
	
	\bibitem{Belli}
	Antonelli F, Campa A, Esposito G, Giardullo P, Belli M, Dini V and et~al 2015
	{\em Radiation Research\/} {\bf 183} 417--431
	\urlprefix\url{https://pubmed.ncbi.nlm.nih.gov/25844944/}
	
	\bibitem{Taucher}
	Jakob B, Splinter J and Taucher-Scholz G 2009 {\em Radiation Research\/} {\bf
		171} 405--418 \urlprefix\url{https://pubmed.ncbi.nlm.nih.gov/19397441/}
	
	\bibitem{Chadwick}
	Chadwick K~H and Leenhouts H~P 1973 {\em Physics in Medicine \& Biology\/} {\bf
		18} 78--87 \urlprefix\url{https://doi.org/10.1088%2F0031-9155%2F18%2F1%2F007}
	
	\bibitem{Tomita}
	Tomita H, Kai M, Kusama T and Aoki Y 1995 {\em Journal of Radiation Research\/}
	{\bf 36} 46--55 \urlprefix\url{https://doi.org/10.1269/jrr.36.46}
	
	\bibitem{Deppman2004}
	Deppman A, Echeimberg J~O, Gouveia A~N, Arruda‑Neto J~D~T, Milian F~M, Added
	N, Camargo M~E, Guzman F, Helene O~A~M, Likhachev V~P, Rodriguez O, Schenberg
	A~C~G, Vanin V and Vicente E~J 2004 {\em Brazilian Journal of Physics\/} {\bf
		34} 958--961 \urlprefix\url{https://doi.org/10.1590/S0103-97332004000500068}
	
	\bibitem{Ramos}
	Ramos-M\'endez J, Garc\'ia-Garc\'ia O, Dom\'inguez-Kondo J, LaVerne J~A,
	Schuemann J, Moreno-Barbosa E and Faddegon B 2022 {\em Physics in Medicine \&
		Biology\/} {\bf 67} 145007
	\urlprefix\url{https://dx.doi.org/10.1088/1361-6560/ac79f9}
	
	\bibitem{Peyrard1}
	Peyrard M and Bishop A~R 1989 {\em Phys. Rev. Lett.\/} {\bf 62}(23) 2755--2758
	\urlprefix\url{https://link.aps.org/doi/10.1103/PhysRevLett.62.2755}
	
	\bibitem{Dauxois1993}
	Dauxois T, Peyrard M and Bishop A~R 1993 {\em Phys. Rev. E\/} {\bf 47}(1)
	684--695 \urlprefix\url{https://link.aps.org/doi/10.1103/PhysRevE.47.684}
	
	\bibitem{RBbook}
	Joiner Michael C and van~der Kogel A~J 2009 {\em Basic Clinical Radiobiology\/}
	4th ed (CRC Press) ISBN 978 0 340 929 667
	
	\bibitem{Scholz}
	Scholz M and Kraft G 1994 {\em Radiation Protection Dosimetry\/} {\bf 52}
	29--33 \urlprefix\url{https://doi.org/10.1093/oxfordjournals.rpd.a082156}
	
	\bibitem{Radford}
	Radford I~R 1985 {\em International Journal of Radiation Biology and Related
		Studies in Physics, Chemistry and Medicine\/} {\bf 48} 45--54
	\urlprefix\url{https://doi.org/10.1080/09553008514551051}
	
	\bibitem{Fu}
	Fu Q, Wang J and Huang T 2018 {\em Journal of radiation research\/} {\bf 1}
	577--582
	\urlprefix\url{https://www.ncbi.nlm.nih.gov/pmc/articles/PMC6151638/}
	
	\bibitem{McMahon}
	McMahon S~J 2018 {\em Physics in Medicine \& Biology\/} {\bf 64} 01TR01
	\urlprefix\url{https://iopscience.iop.org/article/10.1088/1361-6560/aaf26a}
	
	\bibitem{Wang}
	Zheng D, Preuss K, Milano M~T, He X, Gou L, Shi Y, Marples B, Wan R, Yu H, Du H
	and Zhang C 2025 {\em Radiation Oncology\/} {\bf 20} 1--19
	\urlprefix\url{https://ro-journal.biomedcentral.com/articles/10.1186/s13014-025-02626-7}
	
	\bibitem{Lea1942}
	Lea D~E and Catcheside D~G 1942 {\em Journal of Genetics\/} {\bf 44} 216--245
	\urlprefix\url{https://link.springer.com/article/10.1007/BF02982830}
	
	\bibitem{Scalapino}
	Scalapino D~J, Sears M and Ferrell R~A 1972 {\em Phys. Rev. B\/} {\bf 6}(9)
	3409--3416 \urlprefix\url{https://link.aps.org/doi/10.1103/PhysRevB.6.3409}
	
	\bibitem{Krumhans}
	Krumhansl J~A and Schrieffer J~R 1975 {\em Phys. Rev. B\/} {\bf 11}(9)
	3535--3545 \urlprefix\url{https://link.aps.org/doi/10.1103/PhysRevB.11.3535}
	
	\bibitem{Currie}
	Currie J~F, Krumhansl J~A, Bishop A~R and Trullinger S~E 1980 {\em Phys. Rev.
		B\/} {\bf 22}(2) 477--496
	\urlprefix\url{https://link.aps.org/doi/10.1103/PhysRevB.22.477}
	
	\bibitem{Kassis}
	Kassis S, Grondin M and Averill-Bates D~A 2021 {\em Biochimica et Biophysica
		Acta (BBA) - Molecular Cell Research\/} {\bf 1868} 118924 ISSN 0167-4889
	\urlprefix\url{https://www.sciencedirect.com/science/article/pii/S0167488920302822}
	
	\bibitem{Alberts}
	Alberts B, Bray D, Lewis J, Raff M, Roberts K and Watson J 2002 {\em {Molecular
			Biology of the Cell}\/} 4th ed (Garland)
	
	\bibitem{Dikomey}
	Dikomey E and Jung H~W 1991 {\em International journal of radiation biology\/}
	{\bf 59} 815--25 \urlprefix\url{https://doi.org/10.1080/09553009114550711}
	
	\bibitem{AltanBonnet2003}
	Altan-Bonnet G, Libchaber A and Krichevsky O 2003 {\em Phys. Rev. Lett.\/} {\bf
		90}(13) 138101
	\urlprefix\url{https://link.aps.org/doi/10.1103/PhysRevLett.90.138101}
	
	\bibitem{Jose2012}
	Jose D, Weitzel S~E and von Hippel P~H 2012 {\em Proceedings of the National
		Academy of Sciences\/} {\bf 109} 14428--14433
	\urlprefix\url{https://www.pnas.org/doi/abs/10.1073/pnas.1212929109}
	
	\bibitem{DAbramo}
	D’Abramo M, Castellazzi C~L, Orozco M and Amadei A 2013 {\em The Journal of
		Physical Chemistry B\/} {\bf 117} 8697--8704 pMID: 23799235
	\urlprefix\url{https://doi.org/10.1021/jp403369k}
	
	\bibitem{Simon}
	Simon B 2005 {\em Functional Integration and Quantum Physics\/} 2nd ed AMS
	Chelsea Series (Providence, RI: AMS Chelsea Publishing) ISBN 978-0821839469
	
\end{thebibliography}
\providecommand{\newblock}{}

\providecommand{\newblock}{}

\end{document}